\documentclass{article}
\usepackage{arxiv}

\usepackage[utf8]{inputenc} 
\usepackage[T1]{fontenc}    
\usepackage{hyperref}       
\usepackage{url}            
\usepackage{booktabs}       
\usepackage{amsfonts}       
\usepackage{nicefrac}       
\usepackage{microtype}      
\usepackage{lipsum}		
\usepackage{graphicx}
\usepackage{doi}
\usepackage{braket}
\usepackage{amsmath}
\usepackage{appendix}

\title{Two-Step Quantum Search Algorithm for Solving Traveling Salesman Problems}

\author{ Rei Sato$^{1,2}$, Cui Gordon$^{3}$, Kazuhiro Saito$^{1}$, Hideyuki Kawashima$^{3}$, Tetsuro Nikuni$^{2}$, Shohei Watabe$^{4}$\\
	$\rm^1$ KDDI Research, Inc., Fujimino, Ohara 2--1--15, Saitama, 356-8502, Japan\\
 $\rm^2$Department of Physics, Tokyo University of Science, Shinjuku, Tokyo, 162-8601, Japan\\
 $\rm^3$Faculty of Environment and Information Studies, Keio University, Fujisawa, Kanagawa, 252-0882, Japan\\
 $\rm ^4$Faculty of Engineering, Computer Science and Engineering, Shibaura Institute of Technology\\, Toyosu, Tokyo, 135-8548, Japan\\
}

\renewcommand{\shorttitle}{\textit{arXiv} Template}

\hypersetup{
pdftitle={A template for the arxiv style},
pdfsubject={q-bio.NC, q-bio.QM},
pdfauthor={David S.~Hippocampus, Elias D.~Striatum},
pdfkeywords={First keyword, Second keyword, More},
}

\begin{document}
\maketitle

\begin{abstract}
	Quantum search algorithms, such as Grover’s algorithm, are anticipated to efficiently solve constrained combinatorial optimization problems.
	However, applying these algorithms to the traveling salesman problem (TSP) on a quantum circuit presents a significant challenge.  Existing quantum search algorithms for the TSP typically assume that an initial state---an equal superposition of all feasible solutions satisfying the problem’s constraints---is pre-prepared.  The query complexity of preparing this state using brute-force methods scales exponentially with the factorial growth of feasible solutions, creating a significant hurdle in designing quantum circuits for large-scale TSPs.  To address this issue, we propose a two-step quantum search (TSQS) algorithm that employs two sets of operators. In the first step, all the feasible solutions are amplified into their equal superposition state.  In the second step, the optimal solution state is amplified from this superposition state.  The TSQS algorithm demonstrates greater efficiency compared to conventional search algorithms that employ a single oracle operator for finding a solution within the encoded space.  Encoded in the higher-order unconstrained binary optimization (HOBO) representation, our approach significantly reduces the qubit requirements. This enables efficient initial state preparation through a unified circuit design, offering a quadratic speedup in solving the TSP without prior knowledge of feasible solutions.

\end{abstract}


\section{Introduction\label{sec:introduction}}
 
The traveling salesman problem (TSP)~\cite{dantzig1954solution}, which is recognized as NP-hard, stands as a fundamental optimization problem encountered across various engineering fields. 
Quantum algorithms are anticipated to serve as potent tools and have been extensively studied for optimization problems. 
This is owing to their capacity to explore all candidate solutions simultaneously through quantum superposition. 
Leveraging quantum algorithms as solvers for combinatorial optimization dilemmas is expected to yield advantages across a wide range of industries, including portfolio optimization \cite{slate2021quantum}, traffic optimization \cite{neukart2017traffic}, and vehicle routing optimization~\cite{dantzig1959truck}.

Various quantum algorithms have been studied to solve the TSP, offering potential speedups over classical heuristic approaches. These can be broadly categorized into quantum annealing~\cite{PhysRevE.70.057701}, variational methods including variational quantum eigensolvers (VQE)~\cite{matsuo2023enhancing} and quantum approximate optimization algorithms (QAOA)~\cite{ruan2020quantum,salehi2022unconstrained,qian2023comparative,glos2022space}, phase estimation methods~\cite{srinivasan2018efficient,tszyunsi2023quantum}, quantum walk-based methods~\cite{PhysRevResearch.2.023302} and quantum search algorithms~\cite{bang2012quantum,koch2022gaussian,zhu2022realizable,tszyunsi2023quantum} including Grover's algorithm~\cite{grover1996fast}.  
Among these approaches, quantum search algorithms stand out as promising for solving the TSP, offering a quadratic speedup over classical methods~\cite{bang2012quantum, koch2022gaussian, zhu2022realizable}.
Given this potential, we focus on the quantum search algorithm for solving the TSP in this study.

Quantum search algorithms begin by preparing a uniform superposition of all, or a chosen subset of, basis states, subsequently applying followed by the application of a Grover operator~\cite{PhysRevA.105.062453}. The Grover operator comprises oracle operators and the Grover diffusion operator. When solving the TSP with quantum search algorithms, the initial state is typically an equal superposition of all feasible solutions, and the oracle operator serves as the cost oracle~\cite{bang2012quantum,koch2022gaussian, bartschi2020grover}. The cost oracle adjusts the phase of the quantum state based on the TSP tour costs, enabling a quadratic speedup in finding the optimal solution under certain conditions~\cite{bang2012quantum,koch2022gaussian}.

Although many innovative quantum search algorithms have been explored theoretically~\cite{bang2012quantum,koch2022gaussian,tszyunsi2023quantum}, practical challenges arise when constructing circuits for these algorithms to solve the TSP. One such challenge is preparing the initial state~\cite{10313781}, defined as \begin{equation} 
\ket{\psi_0} = \frac{1}{\sqrt{n!}} \sum_i \ket{T_i}, 
    \label{eq:intro_initial_state}
\end{equation} 
where $\ket{T_i}$ represents the feasible solutions to the TSP and $n$ the number of cities. These quantum search algorithms operate by searching within the solution space, and implementing them on a quantum circuit requires preparing all $n!$ states. If a brute-force approach is used for state preparation, the maximum query complexity becomes $\mathcal{O}(n!)$. Although the quantum search algorithm offers a quadratic speedup in exploring the solution space, the overall query complexity including state preparation remains at most $\mathcal{O}(n!)$. This presents a significant obstacle when solving large-scale TSPs. Therefore, efficient preparation of the initial state in Eq.~(\ref{eq:intro_initial_state}) is crucial.

In this study, we propose a two-step quantum search algorithm (TSQS) that enables efficient initial state preparation through a unified circuit design, offering a quadratic speedup in solving TSP instances without prior knowledge of feasible solutions. The proposed circuit architecture incorporates two distinct quantum search processes.

The first step involves performing a quantum search to identify all feasible solutions and generate an equal superposition of these solutions using Grover’s algorithm. The query complexity of preparing the initial state depends on the encoding scheme used for the TSP. Quantum algorithms for the TSP typically utilize one of two encoding methods: Higher-Order Unconstrained Binary Optimization (HOBO) or Quadratic Unconstrained Binary Optimization (QUBO). HOBO employs binary encoding, while QUBO uses one-hot encoding. A key advantage of HOBO is its ability to reduce qubit requirements compared to QUBO, lowering the query complexity from $\mathcal{O}(\sqrt{2^{n^2}/n!})$ to $\mathcal{O}(\sqrt{2^{n\log_2{n}}/n!})$. This reduction brings the query complexity for preparing the equal superposition state below $\mathcal{O}(n!)$. To the best of our knowledge, a detailed quantum circuit design for Grover’s algorithm to prepare all feasible solutions for HOBO-encoded TSP has yet to be explored, particularly given that HOBO encoding is more complex than QUBO encoding.

The second step involves a quantum search to amplify the optimal solution state from the state prepared in the first step. 
In this step, we leverage the quantum circuit used in the first step, which facilitates the construction of a generalized Grover diffusion operator for solving the TSP. Under certain conditions, the query complexity of solving the TSP can be achieved at $\mathcal{O}(\sqrt{n!})$~\cite{bang2012quantum}.

Therefore, the overall query complexity of our algorithm in the HOBO encodling is $ \mathcal{O}(\sqrt{n!})$, which is significantly less than the brute-force method with $\mathcal{O}(n!)$. Our novel framework, based on the proposed two-step circuits, effectively solves the TSP without requiring prior knowledge of the constraints.

The structure of this paper is as follows: In Sec.~\ref{sec:related work}, we briefly review previous studies related to the TSP and quantum search algorithms. In Sec.~\ref{sec:preliminaries}, we describe and formulate the TSP, followed by the introduction of a quantum search algorithm for solving the TSP.  Section~\ref{sec:research problem} explains problem settings.  In Sec.~\ref{sec:propose method}, we present our proposed method. 
Section~\ref{sec:experiments} assesses the performance of our proposed circuits.  Section~\ref{sec:discussion} discusses our results and outlines future research directions. 
Finally, Sec.~\ref{sec:conclusion} summarizes our conclusions.

\section{Related work}
\label{sec:related work}
Quantum algorithms to solving the TSP include a dynamic programming-based algorithm that operates in $\mathcal{O}(1.728^n)$ time~\cite{ambainis2019quantum}, a hybrid quantum algorithm for identifying minimum values in unsorted lists~\cite{durr1996quantum}, and a Grover-based heuristic algorithm that demonstrates quadratic speedup for Gaussian-distributed tour costs~\cite{bang2012quantum}.

Simulation studies have investigated novel oracle operators and qudit states to enhance success probabilities~\cite{koch2022gaussian}, as well as the Grover adaptive search (GAS) algorithm~\cite{Gilliam2021groveradaptive}. Additionally, efforts have been made to design efficient circuits requiring fewer qubits~\cite{zhu2022realizable,sano2023accelerating}.

Other gate-based approaches incorporate phase estimation techniques~\cite{srinivasan2018efficient,tszyunsi2023quantum} and Grover Mixers for Quantum Approximate Optimization Algorithm (GM-QAOA)\cite{bartschi2020grover}. The GM-QAOA employs Grover-like selective phase shift mixing operators, enabling effective searches within the feasible solution space. 

Additionally, a divide-and-conquer quantum search algorithm~\cite{10313781, zhang2020depth} has been proposed to optimize the circuit depth of the Grover’s algorithm by dividing the Grover diffusion operators.  Ref.~\cite{zhang2020depth} proposes a local quantum search algorithm that applies local diffusion operators to the part of whole search space, achieving optimization in certain scenarios compared to the standard Grover’s algorithm. While the diffusion operators are applied locally within search spaces, the size of the oracle remains fixed, encompassing the entire search space to ensure accurate marking of target states.

While these methods have advanced the resolution of the TSP within the quantum information domain, challenges related to initial state preparation and qubit requirements remain.  To address these issues, TSQS algorithm utilizes HOBO encoding, thereby providing a unified circuit design that facilitates both initial state preparation and TSP solving, where the space size of the oracle operator and the Grover diffusion operator is different from each step, compared to Ref.~\cite{zhang2020depth} where the size of both operators are the same.  TSQS algorithm takes multiple steps on a quantum circuit to prepare the solution space, respectively.  This approach offers the potential for a more efficient solution tailored for near-term quantum devices.

\section{Preliminaries}
\label{sec:preliminaries}
\subsection{General encoding of TSP}
\label{sec:TSP}
\begin{figure}
    \centering
    \includegraphics[width=0.8\linewidth]{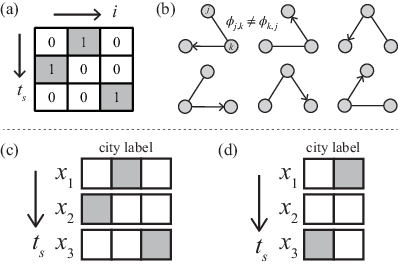}
    \caption{(a) One of feasible solutions $x$ of TSP with $n=3$.  (b) List of all feasible solutions $\ket{T}$, and their tour costs of TSP for $n=3$.  Visualization of a feasible solution of $n=3$ TSP encoded by (a) QUBO and (b) HOBO. The gray and white tiles are equal to $1$ and $0$, respectively}
    \label{fig:general_tsp}
\end{figure}

The TSP is an optimization problem that seeks to determine the tour with minimal cost, where a salesman visits each city exactly once while incurring the least total travel expense. We begin by introducing the general binary encoding of the TSP for $ n $ cities. Let $ x_{t_s,i} $ be a binary variable defined such that $ x_{t_s,i}=1 $ if the $ i $-th city is visited at time $ t_s $, and $ x_{t_s,i}=0 $ otherwise. This encoding method is referred to as the QUBO representation.
We denote the tour cost from city $ i $ to city $ j $ as $ \phi_{ij} $. In our formulation, we consider the asymmetric TSP, where the tour costs are not symmetric; that is, we assume $ \phi_{i,j} \neq \phi_{j,i} $ for two cities $ i $ and $ j $, as illustrated in Fig.~\ref{fig:general_tsp}. Furthermore, we assume that all tour costs are positive, such that $ \phi_{j,k} > 0 $.

The objective function is expressed as follows:
\begin{equation}
  \hat{H}_c(x) = \sum_{i,j=1,i\neq j}^n \phi_{ij} \sum_{t_s=1}^n x_{t_s,i} x_{t_s+1,j}. 
  \label{eq:obj_tsp}
\end{equation}
The TSP is subject to two constraints: precisely one city must be visited at each time step, which can be represented as
\begin{equation}
    \hat{H}_1(x) = \sum_{t_s=1}^{n} \left(1 - \sum_{i=1}^{n} x_{t_s,i}\right)^2=0,
    \label{eq:const1}
\end{equation}
and
\begin{equation}
    \hat{H}_2(x) = \sum_{i=1}^{n} \left(1 - \sum_{t_s=1}^{n} x_{t_s,i}\right)^2=0.
    \label{eq:const2}
\end{equation}
We denote the set of all possible tours that satisfy Eqs.~(\ref{eq:const1}) and (\ref{eq:const2}) as $ \mathcal{T} = \{T_1, T_2, \ldots, T_{n!}\} $, and similarly define the set of all possible tour costs as $ \mathcal{W} = \{W(T_1), W(T_2), \ldots, W(T_{n!})\} $, where $ W(T) $ represents the cost of a tour $ T $. Our goal is to identify $ T_{\min} $ such that $ W(T_{\min}) = \min(W) $, where $ T_{\min} $ is the tour with the minimum cost.

\subsection{Quantum search for solving TSP}
\label{sec:quantum search for solving TSP}


The time evolution of a quantum state in the quantum search algorithm is described by
\begin{equation}
    \ket{\psi(t)} = [\hat{D}\hat{R}]^t \ket{\psi(0)}, 
    \label{eq:rp_time-evolution}
\end{equation}
where $ \ket{\psi(0)} $ is the initial state given by
\begin{equation}
    \ket{\psi(0)} = \frac{1}{\sqrt{n!}} \sum_{i=1}^{n!} \ket{T_i}.
    \label{eq:rp_initial_state}
\end{equation}
Here, $ \hat{R} $ is the cost oracle operator, which provides the tour cost such that
\begin{equation}
    \hat{R} \ket{T_i} = e^{iW(T_i)} \ket{T_i},
    \label{eq:rp_cost_oracle}
\end{equation}
where each cost phase is defined as $ W(T_i) \in \{0, 2\pi\} $, scaled according to the tour costs. The state $ \ket{T_i} $ represents the $ i $-th tour state and is defined as
\begin{equation}
    \ket{T_i} = \ket{x_1} \otimes \ket{x_2} \otimes \cdots \otimes \ket{x_{n}},
\end{equation}
where $ \ket{x_{t_s}} $ corresponds to the $ t_s $-th visited city.

The operator $ \hat{D} $ is the Grover diffusion operator, expressed as
\begin{equation}
    \hat{D} = 2 \ket{\psi(0)} \bra{\psi(0)} - \hat{I}.
    \label{eq:rp_diffuser_D}
\end{equation}
The success probability $ P $ of finding the minimum cost tour state $ \ket{T_{\min}} $ is given by
\begin{equation}
    P = |\braket{T_{\min}|\psi(t_2)}|^2.
    \label{eq:rp_success_probab}
\end{equation}
If the tour costs follow the Gaussian distribution, the optimal query complexity $ t_2 $ is given by
\begin{equation}
    t_2 = \frac{\pi}{4} \sqrt{\frac{n!}{m}},
    \label{eq:bang_t_2}
\end{equation}
where $ m $ represents the number of solutions. This algorithm amplifies the states of both the minimum and maximum cost tours, resulting in $ m=2 $ in this case (see Appendix~\ref{FirstAppendix}).

\section{Research problem}
\label{sec:research problem}

\begin{figure}
    \centering
    \includegraphics[width=80mm]{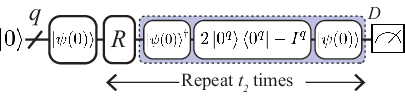}
    \caption{A quantum circuit for the quantum search algorithm used to solve the TSP. }
    \label{fig:research_problem}
\end{figure}

Fig.~\ref{fig:research_problem} illustrates a quantum circuit corresponding to Eq.~(\ref{eq:rp_time-evolution}). Designing this circuit presents several challenges. Specifically, it must generate the state described in Eq.~(\ref{eq:rp_initial_state}). For the TSP with $ n! $ solutions for $ n $ cities, a circuit capable of producing a superposition of these $ n! $ feasible solutions is essential, with a query complexity lower than that of the brute-force method, which is $ \mathcal{O}(n!) $.

Moreover, efficient circuit design for Eq.~(\ref{eq:rp_diffuser_D}) is also challenging, as the Grover diffusion operator depends on the initial state given in Eq.~(\ref{eq:rp_initial_state}). Therefore, the efficient preparation of the Grover diffusion operator $ \hat{D} $ is crucial.

To address these issues, the GM-QAOA algorithm was proposed in Ref.~\cite{bartschi2020grover}, which generates all feasible solutions for the TSP and solves the problem. This approach utilizes the QUBO formulation for the TSP, requiring at least $ n^2 $ qubits for $ n $ cities. In contrast, our approach employs a HOBO formulation and solves the TSP using a two-step Grover algorithm, thereby reducing the number of qubits for encoding space and the circuit depth.

\section{Proposed Method}
\label{sec:propose method}

\begin{figure}[t]
    \centering
    \includegraphics[width=0.5\linewidth]{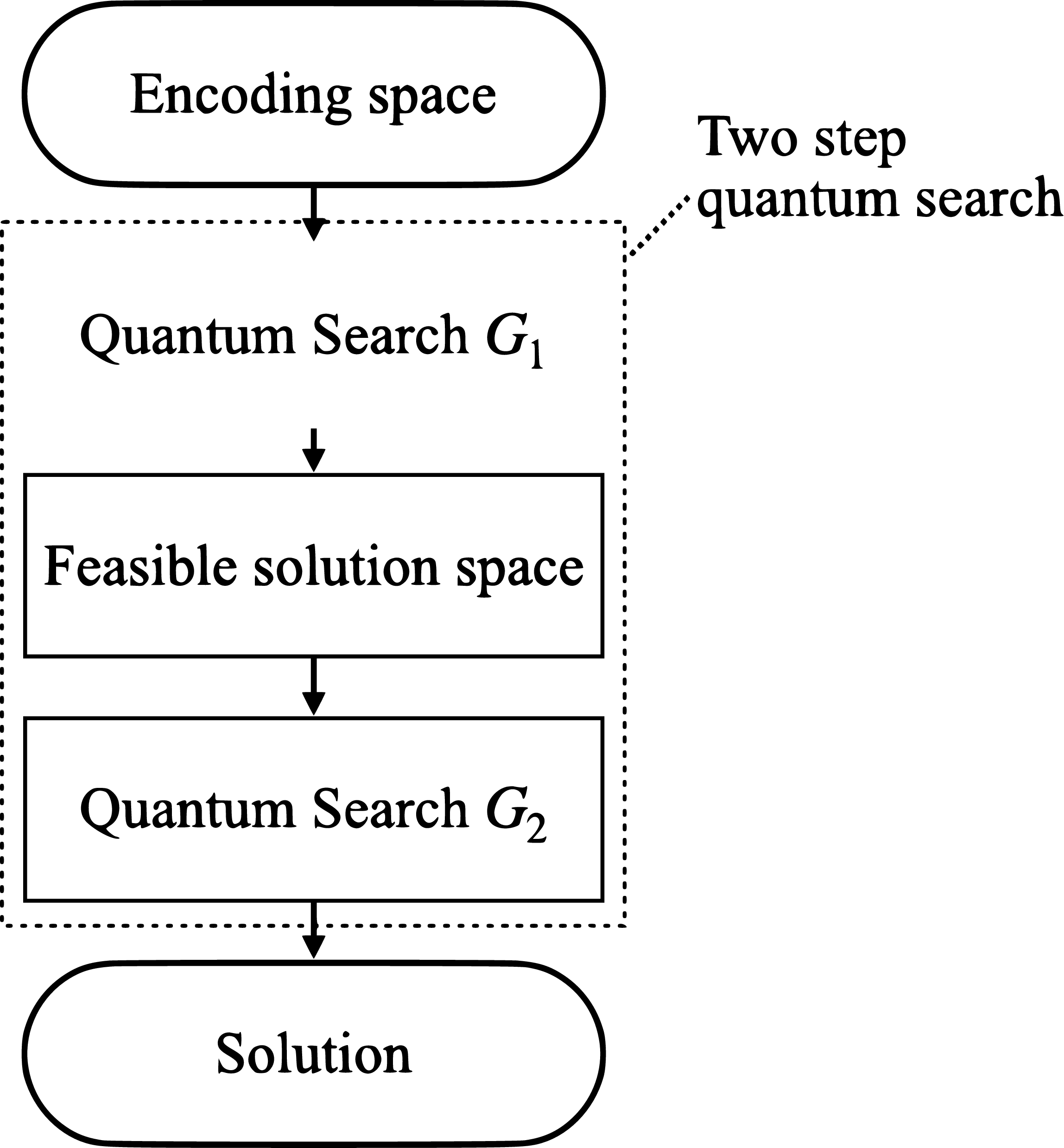}
    \caption{The workflow of the TSQS algorithm.  $\hat{G}_1$ and $\hat{G}_2$ are quantum search operator of preparing all feasible solution and finding optimal solution, respectively.}
    \label{fig:workflow_TSQS}
\end{figure}


Here we propose circuit designs for our TSQS algorithm that overcome the problem of exponentially large query complexity in implementing the initial state~\cite{bang2012quantum}.  Fig.~\ref{fig:workflow_TSQS} shows the workflow of TSQS algorithm.  The encoding space shows all the input data created by the Hadamard transformation.  We employ a two-step quantum search algorithm to find a solution of the combinatorial problem from encoding space.

\subsection{HOBO formulation for TSP}
\label{subsec:propose method-HOBO}

We implement the TSQS algorithm for the TSP using the HOBO formulation described in Ref.~\cite{glos2022space} for encoding space. The HOBO encoding method represents feasible solutions, such as $ \ket{T_i} $, in a binary system, as illustrated in Fig.~\ref{fig:general_tsp}(c)(d). This approach requires $ K = \lceil \log_2{n} \rceil $ qubits for each city, resulting in a total of $ nK $ qubits for encoding HOBO-TSP. In contrast, QUBO encoding necessitates $ n^2 $ qubits due to its one-hot encoding scheme.

For example, Fig.~\ref{fig:general_tsp}(d) illustrates the encoding of a feasible solution where $ 2 $ qubits are needed to encode the cities, such as $ x_1 = 2 = \ket{01} $, $ x_2 = 1 = \ket{00} $, and $ x_3 = 3 = \ket{10} $. When $ 2^K \neq n $, the state $ \ket{11} $ is penalized and not utilized. In the case of $ n=4 $, $ 2^K=n $ holds, meaning all qubits are employed without penalty.

We define the city encoding as 
\begin{align}
 \ket{x_{t_s}} = \ket{x_{t_s,0}, x_{t_s,1}, \ldots, x_{t_s,k}, \ldots, x_{t_s,K-1}} , 
\end{align} 
 where $ x_{t_s,k} $ is the $ k $-th individual qubit associated with encoding city $ x_{t_s} $. Further mathematical details can be found in Ref.~\cite{glos2022space}. A feasible solution state, such as $ \ket{T_i} = \ket{01}\ket{00}\ket{10} $, represents all possible tours using permutations of $ x_{t_s} $ (see Appendix~\ref{FirstAppendix}). The HOBO formulation helps in reducing the query complexity of preparing the equal superposition state of feasible solutions.

\subsection{TSQS algorithm}
\begin{figure*}[t!]
    \centering
    \includegraphics[width=1.0\linewidth]{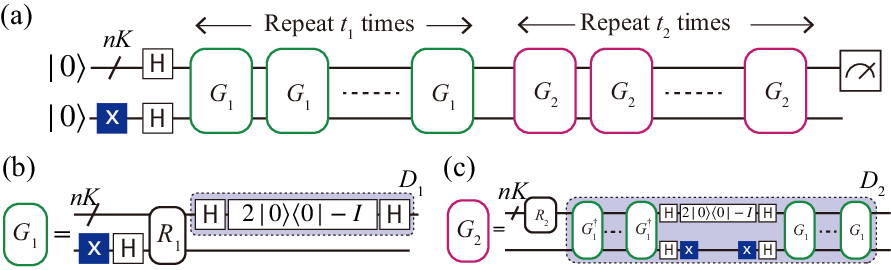}
    \caption{Circuit design of the TSQS algorithm, which prepares the initial state and solves the TSP.}
    \label{fig:TSQS}
\end{figure*}

The time evolution of the TSQS algorithm is given by
\begin{equation}
    \ket{\psi(t_2,t_1)} = \hat{G}_2^{t_2} \hat{G}_1^{t_1} \ket{\psi(0)},
    \label{eq:two_step_state}
\end{equation}
where $ \ket{\psi(0)} = \hat{H}^{nK} \ket{0}^{nK} $, and $ \hat{H} $ is the Hadamard gate. The parameters $ t_1 $ and $ t_2 $ represent the optimal times for the first and second step operations, respectively. 
The operator $ \hat{G}_1 $ is the first step quantum search operator, which prepares an equal superposition state of all feasible solution states for the TSP.
The second step quantum search operator, $ \hat{G}_2 $, finds the optimal solution from all the feasible solutions amplified in the first step (see Fig.~\ref{fig:TSQS} (a)).

The quantum search operator $ \hat{G}_1 $ is a conventional Grover operator, consisting of two unitary operators:
\begin{equation}
    \hat{G}_1 = \hat{D}_1 \hat{R}_1,
\end{equation}
where $ \hat{R}_1 $ is the oracle operator that distinguishes between feasible and non-feasible solution states by marking the solution states with a phase flip:
\begin{equation}
\hat{R}_1 \ket{x} =
\left\{
\begin{array}{rl}
-\ket{x} & \text{if } x = T_i,\\
\ket{x} & \text{if } x \neq T_i,\\
\end{array}
\right.
\label{eq:oracle_R1}
\end{equation}
where $ x $ is an arbitrary binary vector of length $ nK $.

The operator $ \hat{D}_1 $ is the Grover diffusion operator, defined as:
\begin{equation}
    \hat{D}_1 = 2 \ket{\psi(0)}\bra{\psi(0)} - \hat{I}^{nK},
\end{equation}
where $ \hat{I} $ is the identity operator. The Grover diffusion operator facilitates an inversion about the mean (see Fig.~\ref{fig:TSQS} (b)).

The optimal query complexity $ t_1 $ for the first step is given by:
\begin{equation}
    t_1 = \frac{\pi}{4} \sqrt{\frac{2^{nK}}{M}},
    \label{eq:t1}
\end{equation}
where $ M $ is the total number of feasible solutions, i.e., $ M = n! $ for $ n $ cities in the TSP.
In particular, the QUBO encoding requires the optimal query complexity $t_1$ as $\mathcal{O}(\sqrt{2^{n^2}/n!})$, and the HOBO encoding $\mathcal{O}(\sqrt{2^{n\log_2{n}}/n!})$. 

The second quantum search operator, $ \hat{G}_2 $, is used to amplify the optimal solutions for solving the TSP and is given by
\begin{equation}
    \hat{G}_2 = \hat{D}_2\hat{R}_2,
    \label{eq:G2}
\end{equation}
where $ \hat{R}_2 $ is the cost oracle operator, which acts on the state $ \ket{T_i} $ as 
\begin{equation}
    \hat{R}_2\ket{T_i} = e^{-iW(T_i)}\ket{T_i}. 
\end{equation}
Here, the overall cost for a feasible solution $ T_i $ is encoded in the cost phase $ W(T_i) \in \{0, 2\pi\} $, which scales with the tour costs. 
$ \hat{D}_2 $ is the Grover diffusion operator, given by
\begin{equation}
    \hat{D}_2 = 2  \hat{G}_1^{t_1} \ket{\psi(0)}\bra{\psi(0)} \hat{G}_1^{t_1\dagger} - \hat{I},
    \label{eq:D2_1}
\end{equation}
where $ \hat{G}_1^{t_1} \ket{\psi(0)} $ provides the equal superposition of all feasible solution states from Eq.~(\ref{eq:rp_initial_state}) generated by the first quantum search step, acting on $ \ket{0}^{nK} $ as 
\begin{equation}
    \hat{G}_1^{t_1} \ket{\psi(0)} \simeq \frac{1}{\sqrt{n!}}\sum_{i}\ket{T_i}. 
    \label{eq:D2_2}
\end{equation}
We design the Grover diffusion operator $ \hat{D}_2 $ based on the first quantum search operator. 
Fig.~\ref{fig:TSQS} (c) provides a detailed circuit for $ \hat{D}_2 $, which acts only on the basis states corresponding to all the feasible solutions prepared during the first quantum search step. 
The success probability for finding the minimum-cost tour state $ \ket{T_{\min}} $ is given by
\begin{equation}
P(T_{\min},t_1,t_2) = |\braket{T_{\min}|\psi(t_2,t_1)}|^2,
\label{eq:success_prob}
\end{equation}
where $ \ket{\psi(t_2,t_1)} $ is given by Eq.~(\ref{eq:two_step_state}).  

If the cost oracle $ \hat{R}_2 $ follows a Gaussian distribution, the optimal time $ t_2 $ can be estimated as 
\begin{equation}
t_2 = \frac{\pi}{4}\sqrt{ \frac{n!}{2} }.
\label{eq:t2}
\end{equation}
This value of $ t_2 $ matches Eq.~(\ref{eq:bang_t_2}) for $ m=2 $, where the present algorithm amplifies the states of both the minimum and maximum cost tours. 
The total query complexity of the TSQS algorithm is thus $ t = t_1 + t_2 $.

\section{Circuit design}
\label{sec:experiments}
\begin{figure}[tp]
    \centering
    \includegraphics[width=1.0\linewidth]{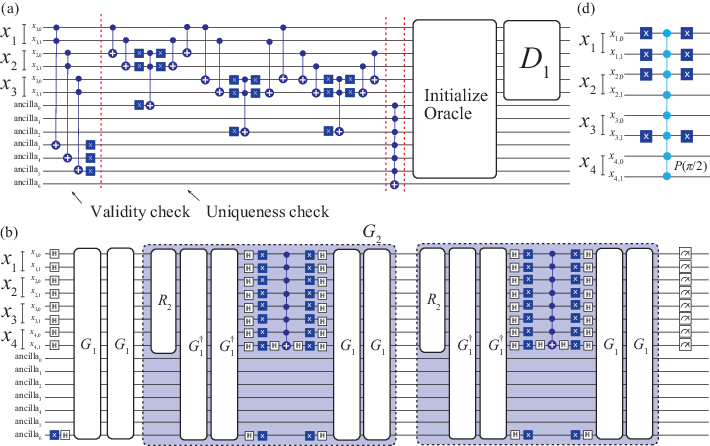}
    \caption{The circuits of TSQS algorithm.  (a) Grover operator circuit of the first quantum search $\hat{G}_1$ in the TSP for $n=3$.  (b) The circuit of the TSQS algorithm for solving the TSP for $n=4$. (c) One of the cost oracles constructed by multi-phase gate $W(T_1)=\pi/2$ for the tour $\ket{00011011}$.}
    \label{fig:two-step_circuit}
\end{figure}

We address the specific circuit structures of the TSQS algorithm by extending the approach presented in Ref.~\cite{bang2012quantum}. 
As examples, we examine TSP problems, as discussed in Sec.~\ref{sec:quantum search for solving TSP}. 
The minimum and maximum cost tour states correspond to the minimum and maximum route costs of $ \pi/2 $ and $ 3\pi/2 $, respectively, while the intermediate states are randomly generated according to a Gaussian distribution, as described in Sec.~\ref{sec:quantum search for solving TSP}. 
Because of the computational capability, we conducted the TSP analysis for $ n=3 $ and $ n=4 $ cities. 
Additionally, we compare our circuit design with that of the GM-QAOA approach~\cite{bartschi2020grover}.

\subsection{Circuit design of the first-step quantum search}
We require two sub-oracles: one for validity checking and the other for ensuring uniqueness in the encoding of the HOBO-TSP, as described in subsection~\ref{subsec:propose method-HOBO}.

The validity check addresses a potential issue in the quantum encoding of cities, which arises in the sub-oracle. Specifically, when qubits are used to binary encode each city, and the number of cities $n$ is not equal to $2^K$ in the HOBO formulation, non-existent cities may be erroneously encoded. For example, with $n=3$ cities, using two qubits for the city representation results in the undesired state $ \ket{11} $, which corresponds to a non-existent city. In contrast, this issue does not arise for $n=4$ cities, where $n=2^K$. This problem occurs only when the number of possible states represented by the qubits, $2^K$, exceeds the actual number of cities, $n$, in the TSP.

To resolve this issue, we employ multi-controlled X gates, also referred to as MCXGate in Qiskit (hereafter referred to as MCX) to filter out invalid binary representations.  In the case of $n=3$ cities, we prevent the state $ \ket{11} $ by using the circuit shown in Fig.~\ref{fig:two-step_circuit}(a). In Fig.~\ref{fig:two-step_circuit}(b), there is no invalid state, as $n=2^K$, and thus, the validity check oracle is not required.

The uniqueness check ensures that each city in the tour is visited exactly once. This is achieved by implementing a function that compares pairs of cities and outputs 0 if they are the same and 1 if they are different. The function is applied to all city pairs, and the state is marked as a valid solution only if all the pairs return 1. The oracle function is defined as:
\begin{equation}
f(x_{t_s}, x_{t'_s}) =
\left\{
    \begin{array}{rll}
    0, & \text{if } x_{t_s} = x_{t'_s}, \\
    1, & \text{if } x_{t_s} \neq x_{t'_s}.  \\
    \end{array}
\right.
\end{equation} 
The sub-oracle ensures $x_{t_s} \neq x_{t'_s}$ by verifying that for all index $k$, there exists at least one $k$ value such that $(x_{t_s,k} \neq x_{t'_s,k})$.  
This is achieved using CNOT gates also referred to as CNOT gates and X (not) gates. 
For each city pair, we apply a CNOT gate with $x_{t_s,k}$ as the control bit and $x_{t'_s,k}$ as the target bit. We then check whether at least one of the target bits is $1$ using an OR gate constructed from CX and X gates. If the OR gate returns $f = 0$, it indicates that $x_{t_s} = x_{t'_s}$; if it returns $f = 1$, then $x_{t_s} \neq x_{t'_s}$. To restore the city qubit to its original state for future use, we apply the CNOT operations again. For example, Fig.~\ref{fig:two-step_circuit}(a)(b) illustrates the circuit pattern, showing a CNOT applied to $x_{1,0}$ and $x_{2,0}$, and another CNOT applied to $x_{1,1}$ and $x_{2,1}$. We implement this checking pattern for all city pairs and utilize an MCT (multi-controlled Toffoli) gate to flip the phase of the state only if all OR gates return a value of $1$, ensuring the oracle condition is met.

The total number of qubits required for the circuit is:
\begin{align}
nK + a_{\text{valid}} + a_{\text{unique}} + 1,
\end{align}
where $ nK $ represents the qubits used for TSP encoding. The ancilla qubits for the validity check are $ a_{\text{valid}} = (2^K - n)n $, and for the uniqueness check $ a_{\text{unique}} = \sum_{i=1}^{n-1} i $. The final ``$+1$'' represents the ancilla qubit used to mark the solution state.

\subsection{Circuit design of the second-step quantum search}

Fig.~\ref{fig:two-step_circuit}(b) depicts an actual circuit design based on Fig.~\ref{fig:TSQS}, illustrating the TSQS algorithm for the TSP with $n=4$. 
First, we apply Hadamard gates to the $nK$ qubits. Next, we implement the first quantum search operator $\hat{G}_1$ for the optimal time, effectively eliminating the states of infeasible solutions and generating a superposition state of feasible solutions as indicated by Eq.~(\ref{eq:rp_initial_state}). Subsequently, we employ the cost oracle as described in Eq.~(\ref{eq:rp_cost_oracle}) to the state, following Ref~\cite{bang2012quantum}. 
For instance, Fig.~\ref{fig:two-step_circuit}(c) illustrates one of the cost oracles constructed using multi-phase gates and the X-gate for the TSP tour $\ket{00011011}$, the cost of which is $W(00011011) = \pi/2$. We construct the cost oracle for all feasible solution states. 

After constructing the cost oracle, we apply the Grover diffusion operator $\hat{D}_2$, which acts solely on the solution space of feasible solutions, as shown in Fig.~\ref{fig:TSQS}(c). The operator $\hat{D}_2$ can be represented using the first Grover operator $\hat{G}_1$, as indicated in Eq.~(\ref{eq:D2_1}). 
The optimal number of operations for $\hat{G}_1$ and $\hat{G}_2$ are determined based on the query complexity outlined in Eqs.~(\ref{eq:t1}) and (\ref{eq:t2}), respectively. Detailed information about each parameter is presented in Table I for the TSP with $n=3$ and $n=4$.

\section{Simulation result}
\label{sec:simulation result}
\begin{figure}[tb]
    \centering
    \includegraphics[width=0.5\linewidth]{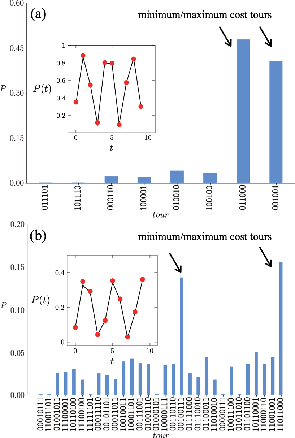}
    \caption{The simulation results of TSQS algorithm.  (a) The histogram of the success probability for $n=3$ TSP.  (b) The histogram of the success probability for $n=4$ TSP.   The insets show the success probability of the TSQS (dots) with the numerical simulation (solid lines).  The success probability includes the probability of minimum and maximum cost tours.}
    \label{fig:two-step_result}
\end{figure}

\begin{table*}[htbp]
    \centering
    \caption{The circuit evaluation among TSQS, SSQS, and GM-QAOA for solving TSP with $n=3,4$ cities. Iteration means iteration for classical optimizer of QAOA.  Shots means number of shots for classical optimizer.  The numbers in parentheses represent the total success probability of the min/max cost tours. $p$ is number of layers of GM-QAOA.}
    \scalebox{0.75}{
    \begin{tabular}{|c|cccc|cccc|}
    \hline
         & \multicolumn{4}{c|}{$n=3$} & \multicolumn{4}{c|}{$n=4$} \\
    & TSQS & SSQS & GM-QAOA & GM-QAOA & TSQS & SSQS & GM-QAOA & GM-QAOA\\
    &  &  & $p=1$ & $p=2$ &  &  & $p=1$ & $p=2$\\
    & (HOBO) & (HOBO) & (QUBO) & (QUBO) & (HOBO) & (HOBO) & (QUBO) & (QUBO)\\\hline\hline
    iteration& -- & -- & $50$ & $50$ & -- & -- & $50$ & $50$ \\
    shots & -- & -- & $100$ & $100$ & -- & -- & $100$ & $100$ \\
    $t_1$& $2$ & -- & -- & -- & $2$ & -- & -- & -- \\
    $t_2$& $1$ & -- & -- & -- & $2$ & -- & -- & -- \\
    total complexity& $3$ & $4$ & $50$ & $50$ & $4$ & $9$ & $50$ & $50$ \\\hline
    width& $13$ & $13$ & $9$ & $9$ & $15$ & $15$ & $16$ & $16$ \\
    depth& $4636$ & $52522$ & $3116$ & $5581$ & $43211$ & $156989$ & $167890$ & $337937$ \\\hline
    $P$& $0.42..(0.97..)$ & $0.26..(0.47..)$ & $0.48\pm0.02$ & $0.58\pm0.19$ & $0.14..(0.31..)$ & $0.22..(0.43..)$ & $0.052\pm 0.026$ & $0.096\pm0.023$ \\\hline
    \end{tabular}
    }
    \label{tab:circuit_evalulation}
\end{table*}

We evaluate the accuracy of the TSQS algorithm and perform benchmarks on circuit width, depth, and the query complexity of preparing the initial state and solving the TSP. The evaluations were conducted using the Qiskit simulator~\cite{Qiskit} within an IBM quantum system. 
We compare the query complexity of the brute-force method, $\mathcal{O}(n!)$, SSQS algorithm, GM-QAOA and TSQS algorithm.   See the Appendix~\ref{SecondAppendix} and \ref{ThirdAppendix} for the simulation results of GM-QAOA and SSQS.

The software versions of Qiskit employed in our experiments are: qiskit-terra:0.21.1, qiskit-aer:0.10.4, qiskit-ignis:0.7.1, qiskit-ibmq-provider:0.19.2, and qiskit:0.37.1. 
For the numerical environment, we fix the seed number as \texttt{seed\_simulator = 42} and \texttt{seed\_transpiler = 42}, and the shot number for measuring as \texttt{shots = 1024}. 

\subsection{Performance evaluation}

Fig.~\ref{fig:two-step_result} illustrates the results of the TSP for $n=3$ and $n=4$ using our TSQS. To verify the circuit operation, we plot the time-dependent success probabilities based on numerical calculations performed in the Julia programming language for matrix computations (solid lines in the inset) and values simulated by the circuit (red dots in the inset).  The time-dependent success probability includes both minimum and maximum cost states.  For both $n=3$ and $n=4$ cities, the numerical results from the matrix computations and the circuit simulations are nearly identical, confirming the proposed circuit's correct operation.

However, in the case of circuit simulations, we observe negligibly small probabilities for non-constrained solutions such as $101110$ and $011101$ for $n=3$, and $11101000$, $10110011$, $10111111$, etc., for $n=4$, as shown in Figs.~\ref{fig:two-step_result} (a) and (b). These occurrences are due to minor errors in the first step of the Grover search.  While the Grover search certainly amplifies feasible solutions, non-feasible solutions may also be observed due to noise. 
Consequently, in the second step of the quantum search, non-feasible solutions are incorporated into the Grover diffusion operator with negligibly small weights. The proportion of non-feasible solutions is significantly smaller compared to the success probability of feasible solutions, ensuring that the circuit operates effectively without a substantial loss of success probability over time, as shown in the insets of Figs.~\ref{fig:two-step_result}(a) and (b).

In addition to these minor issues, the proposed algorithm amplifies both minimum- and maximum-cost solutions due to the phase structure of the cost oracle, as shown in Ref.~\cite{bang2012quantum}.  This characteristic halves the success probability for the minimum-cost solution alone because the final superposition state includes both solutions with similar amplitudes.  However, distinguishing and discarding the maximum-cost solution can be achieved through a simple post-processing of the tour costs, which involves minimal computational overhead $\mathcal{O}(1)$.  The success probability of TSQS reported in Table~I and the insets of Fig.~\ref{fig:two-step_result} reflect the success probabilities, including both minimum and maximum cost states as success probability, where the combined probability for both states remains high, ensuring the robustness of the algorithm.

\begin{figure}[tb]
    \centering
    \includegraphics[width=1.0\linewidth]{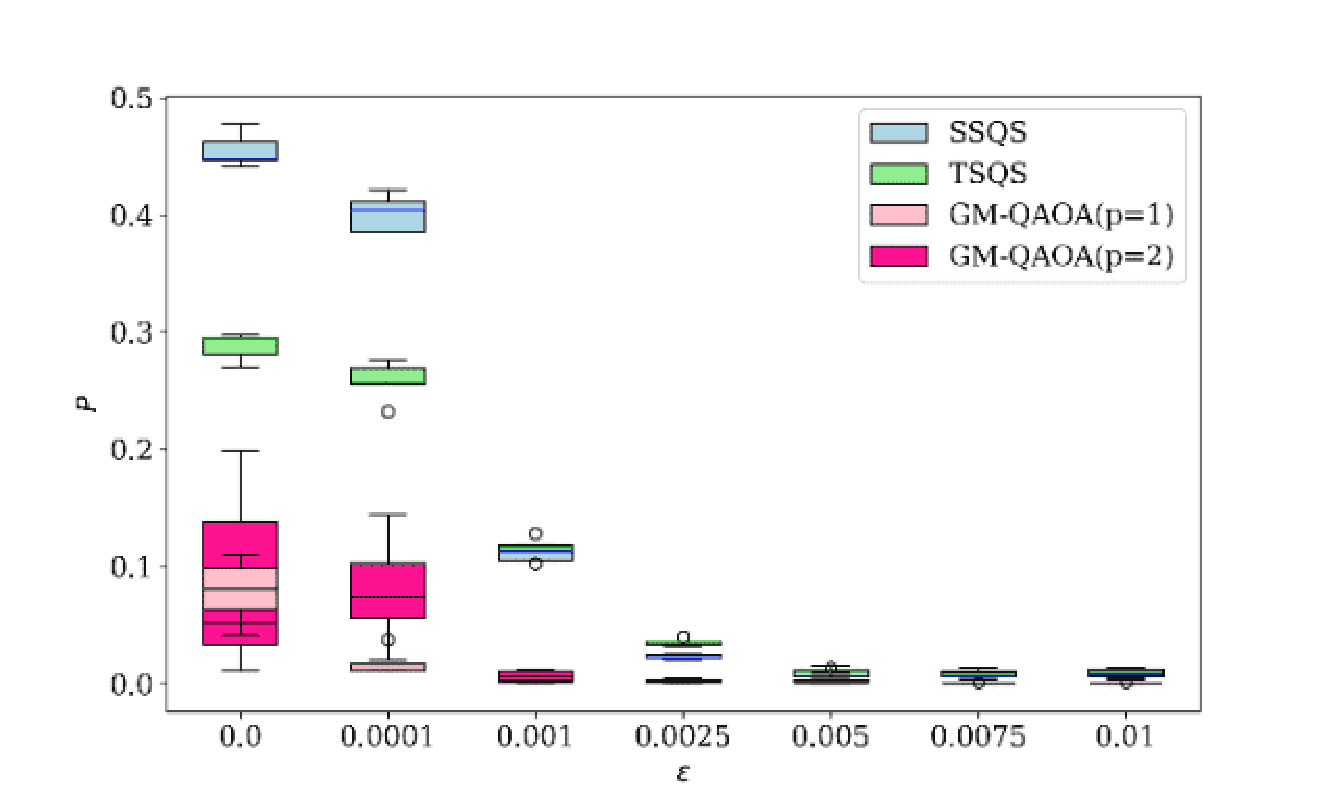}
    \caption{Box plots of test accuracy on $n=4$ TSP with depolarizing noise.  $\epsilon$ is noise level. 
 Each box contains $10$ repeated experiments. 
 The top horizontal line of the box is called the maximum value, the bottom horizontal line is called the minimum value, and the middle line is called the median.  The white circular points represent outliers.}
    \label{fig:noisy_result}
\end{figure}

We also investigate the performance of TSQS, SSQS and GM-QAOA in noisy environment.  We employ the depolarizing channel~\cite{nielsen2010quantum}.  The depolarizing channel $\mathcal{E}$ acting on $n$ qubits described by a density matrix $\rho$ is defined as  
\begin{equation}
    \mathcal{E}(\rho) = \frac{\epsilon \hat{I} }{2^n} + (1 - \epsilon)\rho,
\end{equation}
where $\epsilon$ is the depolarizing noise strength and $\hat{I}$ denotes the identity operator. 

For the quantum circuit implementation in Qiskit considering this depolarization model, we applied the depolarization noise model to single-qubit gates, including the Hadamard gate H, Pauli gate X, Y, Z, and unitary gates u1, u2, and u3, which are parameterized single-qubit rotation gates in Qiskit, as well as the two-qubit CNOT gate~\cite{Qiskit}.  The noise parameter $\epsilon$ was set to $0.0$, $0.0001$, $0.001$, $0.0025$, $0.005$, $0.0075$, $0.01$, $0.02$, and $0.03$. 
 Fig.~\ref{fig:noisy_result} shows the success probability results for the minimum and maximum tours at different noise levels $\epsilon$. When $\epsilon$ is small, SSQS performs better, but as $\epsilon$ increases, TSQS shows slightly better performance. Considering this along with the results in Table.~\ref{tab:circuit_evalulation}, TSQS maybe be well-suited for near-term noisy quantum devices since it has a shallower circuit and lower query complexity.

\subsection{Circuit evaluation}
Table~\ref{tab:circuit_evalulation} presents a circuit evaluation for the TSQS, GM-QAOA~\cite{bartschi2020grover} and SSQS algorithm.  The examples of the simulation results for GM-QAOA and SSQS are provided in Appendix~\ref{SecondAppendix} and Appendix~\ref{ThirdAppendix}, respectively.  

\subsubsection{Complexity}
\begin{figure}[tb]
    \centering
    \includegraphics[width=0.5\linewidth]{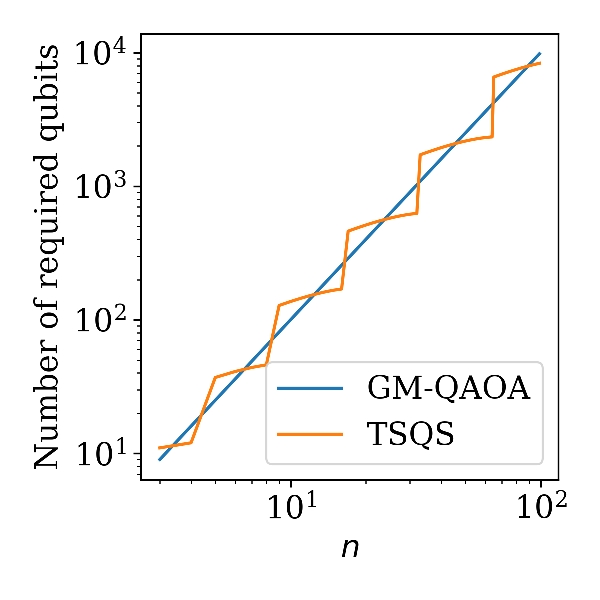}
    \caption{Comparison of the total number of qubits required to implemente GM-QAOA and TSQS.}
    \label{fig:width_TSQSvsQAOA}
\end{figure}

We first evaluated the required number of qubits.  As shown in Fig.~\ref{fig:width_TSQSvsQAOA}, there is no significant average difference in number of required qubits between the TSQS algorithm and GM-QAOA, the TSQS algorithm can be implemented with fewer qubits than GM-QAOA in certain regions, and vice versa.  

On the other hand, it is hard to evaluate the actual depth of each circuit for $n>4$ due to the memory limitation for simulation.  Instead, we estimate that the gate complexity of the oracle operator $\hat{R_1}$ used to prepare the initial state can be implemented with $\mathcal{O}(n^2 \log^2 n)$ gates. For the validity check, the total number of gates required is $(2^K - n)n + n$ for $n$ cities, where the first term represents the number of MCX gates and the second term accounts for the number of X gates. For the uniqueness check, we have $\{K^2 + (2K + 1) + 1\} \times \sum_{i=1}^{n-1} i$, where $K^2$, $(2K + 1)$, and $1$ represent the number of CNOT gates, X gates, and MCX gates, respectively. 
Thus, the total number of gates can be expressed as 
 $\left\{(2^K-n)n + n\right\} + \{K^2+(2K+1)+1\}\times\sum_{i=1}^{n-1}i =
\mathcal{O}(n^2) + \mathcal{O}(\log^2 n) \times \mathcal{O}(n^2) \sim \mathcal{O}(n^2 \log^2 n)$. 
Although the gate complexity of $\hat{R}_1$ is larger than the $\mathcal{O}(n^2)$ of $\hat{U}_s$ in GM-QAOA~\cite{bartschi2020grover}, the total depth of the TSQS circuit is shallower than that of GM-QAOA and SSQS since GM-QAOA requires the MCX gate in the Grover mixer operator scales with larger $n$ and also SSQS requires an additional oracle operator to remove the unsatisfied tour state.

\subsubsection{Query complexity}
Earlier TSP algorithms utilizing quantum search~\cite{bang2012quantum} achieved a query complexity of $\mathcal{O}(\sqrt{n!})$ under the assumption that an equal superposition of all feasible solutions had already been prepared. However, the method for preparing such a state was not explicitly established in~\cite{bang2012quantum}. If a brute-force approach is employed to construct this superposition, the query complexity becomes $\mathcal{O}(n!)$, which nullifies the quadratic speedup gained by the quantum search. In contrast, our TSQS algorithm is significantly faster since it can optimally prepare the superposition of feasible solutions (see Table~\ref{tab:time-complexity}). For example, for TSP instances with $n = 3$ and $n = 4$, the superposition can be prepared with $t_1 = 2$ and $t_2 = 1, 2$, respectively—remarkably smaller than $n!$.
Indeed, in the TSQS algorithm, the first step constructs an equal superposition state of all feasible solutions for the TSP in the encoding space of size $2^{nK}$, where $n$ represents the number of cities and $K$ is the encoding size per city. The optimal query complexity of the first quantum search step is $t_1 = \mathcal{O}(\sqrt{2^{nK}/n!})$. The second step then amplifies the minimum-cost tour from the set of feasible solutions, which has a size of $n!$. The optimal query complexity for this second quantum search step is $t_2 = \mathcal{O}(\sqrt{n!})$.
In the case of QUBO encoding, where $K = n$, the overall query complexity of the TSQS algorithm, $t_\mathrm{TSQS} = t_1 + t_2$, is dominated by the first step, leading to a total complexity of $t_\mathrm{TSQS} = \mathcal{O}(\sqrt{2^{n^2}/n!})$. On the other hand, for the HOBO encoding, where $K = \lceil \log_2 n \rceil$, the second step dominates the query complexity, yielding $t_\mathrm{TSQS} = \mathcal{O}(\sqrt{n!})$. Therefore, the QUBO encoding’s requirement to prepare the equal superposition of all feasible solutions leads to a less efficient query complexity compared to the HOBO encoding.
Alternatively, the single-step quantum search (SSQS) algorithm evolves the system as $\ket{\psi(t)} = (\hat{D}\hat{R})^{t} \ket{\psi(0)}$, where $\hat{R}$ is the oracle operator marking the minimum-cost tour state $\ket{T_{\min}}$, and $\hat{D}$ is the diffusion operator acting on an encoding space of dimension $2^{nK}$. In this case, the optimal query complexity is $t_\mathrm{SSQS} = \mathcal{O}(\sqrt{2^{nK}})$. Consequently, the SSQS algorithm exhibits a longer query complexity compared to the TSQS algorithm, both for the QUBO encoding ($t_\mathrm{TSQS} = \mathcal{O}(\sqrt{2^{n^2}/n!})$) and the HOBO encoding ($t_\mathrm{TSQS} = \mathcal{O}(\sqrt{n!})$). This advantage is due to the fact that SSQS searches for the minimum-cost tour across the entire encoding space, whereas TSQS restricts the search to the smaller space of feasible solutions, which is extracted from the encoding space.  As shown in Table~\ref{tab:circuit_evalulation}, the query complexity required to amplify the minimum-cost tour state from the entire encoding is larger for the SSQS algorithm compared to TSQS. In addition to the increase in query complexity, SSQS has a significantly deeper circuit depth than TSQS due to the complexity of oracle design (see Appendix~\ref{ThirdAppendix} for detail information). This indicates that TSQS is advantageous in these aspects.


Furthermore, the TSQS algorithm outperforms the GM-QAOA algorithm in terms of query complexity for solving the TSP.  This is because GM-QAOA requires multiple iterations of classical optimization, typically ranging from 10 to 20 iterations for TSPs with $n = 3$ and $n = 4$ cities, respectively.  Notably, the number of classical iterations in GM-QAOA grows as the number of QAOA layers, $p$, increases.  Furthermore, the TSQS algorithm outperforms the GM-QAOA algorithm in terms of query complexity for solving the TSP.  This is because GM-QAOA requires multiple iterations of classical optimization, typically ranging from 10 to 20 iterations for TSPs with $n = 3$ and $n = 4$ cities, respectively.  Notably, the number of classical iterations in GM-QAOA grows as the number of QAOA layers, $p$, increases.  However, compared to the TSQS algorithm, the advantages of GM-QAOA are that, aside from the computational cost of the optimization process, the success probability of finding the minimum cost tour increases as $p$ increases, and it can solve a variety of cost tour distributions, not limited to Gaussian distributions.

\begin{table*}[tb]
  \centering
  \caption{Query complexity of SSQS and TSQS algorithms for QUBO and HOBO encodings. TSQS with brute-force preparation is also shown. }
  \scalebox{0.85}{
  \begin{tabular}{|c|cc|ccc|}\hline
  & SSQS (QUBO) & SSQS (HOBO) & TSQS (Brute-force) & TSQS (QUBO) & TSQS (HOBO)
  \\ \hline \hline
  $1^\mathrm{st}$-step (prepare feasible solutions) & - & - & $\mathcal{O}(n!)$ & $\mathcal{O}(\sqrt{2^{n^2}/n!})$ & $\mathcal{O}(\sqrt{2^{n\log_2 n}/n!})$
  \\
  $2^\mathrm{nd}$-step (find min.-cost tour) & - & - & $\mathcal{O}(\sqrt{n!})$ & $\mathcal{O}(\sqrt{n!})$ & $\mathcal{O}(\sqrt{n!})$
  \\ \hline
  total & $\mathcal{O}(\sqrt{2^{n^2}})$ & $\mathcal{O}(\sqrt{2^{n\log_2 n}})$ & $\mathcal{O}(n!)$ & $\mathcal{O}(\sqrt{2^{n^2}/n!})$ & $\mathcal{O}(\sqrt{n!})$\\\hline
  \end{tabular}
  }
  \label{tab:time-complexity}
\end{table*}

\section{Discussion}
\label{sec:discussion}
By using the TSQS algorithm, it is possible to prepare an equal superposition of all feasible solutions with a query complexity lower than $\mathcal{O}(n!)$.  Furthermore, a circuit can be constructed to amplify the tour states corresponding to the minimum and maximum costs, achieving quadratic speedup, especially when the tour costs follow a Gaussian distribution. 

In the comparison of TSQS, SSQS, and GM-QAOA, the success probability is expected to favor TSQS and SSQS as $n$ increases.  Additionally, TSQS almost demonstrated the shallowest circuit depth among the three approaches. Furthermore, in a noisy environment, TSQS exhibited a slight advantage as noise levels increased. On the other hand, there was no significant difference in the number of qubits required to implement TSQS (or SSQS) and GM-QAOA.  However, the number of qubits in TSQS can be reduced by efficiently utilizing ancillary qubits. For instance, in the validity check process, an ancillary qubit is assigned for each city, but there remains room for improvement, such as reducing this to a single ancillary qubit.  

Through this analysis, TSQS and GM-QAOA each have their strengths and weaknesses. TSQS adopts the cost-oracle design from Ref.~\cite{bang2012quantum} because it guarantees quadratic speedup of query complexity for large-scale problems. However, this approach amplifies both the minimum and maximum costs simultaneously, reducing the success probability by half. In contrast, GM-QAOA provides flexible and adaptive approaches capable of amplifying only the minimum cost. However, the large encoding space, which includes infeasible solutions, and the lack of quadratic speedup compared to TSQS, result in increased computational complexity.

As an alternative approach, TSQS could be enhanced by preparing a superposition of feasible solutions with Grover's algorithm in the first step, followed by applying GAS~\cite{Gilliam2021groveradaptive} in the second step to search the reduced solution space. This could enable faster identification of the optimal solution.  This method is expected to dynamically adjust the oracle threshold fewer times than conventional GAS during the search process, potentially allowing for more efficient amplification of minimum-cost states compared to TSQS.

Additionally, general circuit improvement methods such as divide-and-conquer techniques~\cite{qd_Wu_Xi,satoh2020subdivided,zhang2020depth,zhang2021implementation,park2023quantum} may prove beneficial. The depth of the circuit increases significantly with the number of cities, making it challenging to implement large-scale TSPs on current quantum computing systems.  Therefore, advancements in creating shallower circuits are essential.  In practice, TSQS requires $n!$ MCX gates for constructing the cost oracle circuit in the second step. To facilitate circuit implementations with fewer than $n!$ embedded computations, it may be advantageous to decompose the cost oracle into sub-oracles and reuse them through a divide-and-conquer approach.

\section{Conclusions}
\label{sec:conclusion}
We proposed and verified a Two-Step Quantum Search (TSQS) algorithm, along with its circuit construction, which is capable of preparing an equal superposition of all feasible solutions and solving the Traveling Salesman Problem (TSP) on a unified quantum circuit. The TSQS prepares the initial state with a query complexity that is less than that of the brute-force method, $\mathcal{O}(n!)$, and amplifies the tour state corresponding to the minimum cost of the TSP. We tested the proposed method for TSP instances with $n=3$ and $n=4$ cities. Our approach successfully reduces the query complexity for solving the TSP to less than $\mathcal{O}(n!)$; however, it presents challenges related to circuit depth. Developing methods to implement our circuit with shallower architectures remains a future challenge.

\appendices
\section{dataset and Numerical simulation of quantum search}
\label{FirstAppendix}
\begin{table}[htbp]
    \centering
    \caption{Dataset of tour cost for $n=3,4$ TSPs.}
    \begin{tabular}{|cll|}\hline
    $\phi_{i,j}$& $n=3$ & $n=4$ \\\hline\hline
    $\phi_{0,1}$& $1.066..$ & $0.523..$\\
    $\phi_{1,0}$& $2.818..$ & $1.047..$\\
    $\phi_{0,2}$& $0.866..$ & $1.047..$\\
    $\phi_{2,0}$& $2.434..$ & $2.094..$\\
    $\phi_{1,2}$& $0.503..$ & $0.523..$\\
    $\phi_{2,1}$& $1.893..$ & $1.047..$\\
    $\phi_{1,3}$&  & $1.047..$\\
    $\phi_{3,1}$&  & $1.047..$\\
    $\phi_{2,3}$&  & $0.523..$\\
    $\phi_{3,2}$&  & $1.396..$\\\hline
    \end{tabular}
    \label{tab:dataset}
\end{table}

\begin{figure}[tb]
    \centering
    \includegraphics[width=0.5\linewidth]{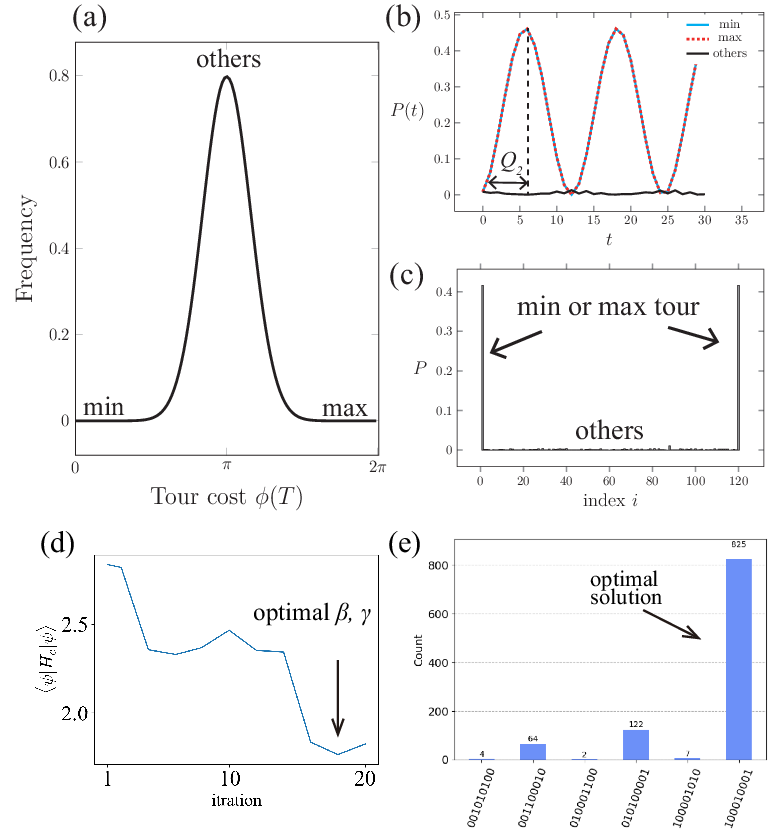}
    \caption{(a)-(c)Numerical simulation of the quantum search for $n=5$ TSP which cost follows Gaussian distribution where $\mu=\pi$ and $\sigma=0.5$. (d)(e) The simulation results of GM-QAOA algorithm for $n=3$ cities.  (a) The Gaussian distribution.  (b) Time-dependence success probability $P(t)$.  The min/max represents minimum/maximum cost tours. (c) The histogram of success probability of all feasible solutions, i.e., $5!=120$ for optimal query complexity $t_2$. 
 (d) Optimization expectation value $\braket{\psi(\beta,\gamma)|\hat{H}_c|\psi(\beta,\gamma)}$ for classical iteration.  (e) The histogram of success probability for $n=3$ TSP with optimized parameters $\beta,\gamma$ and $p=2$.}
    \label{fig:Grover_GM-QAOA}
\end{figure}

The tour costs are assumed to be generated from the Gaussian distribution given by 
\begin{equation}
    f(x) = \frac{1}{\sqrt{2 \pi \sigma^2}} \exp{\left[-\frac{(x - \mu)^2}{2 \sigma^2}\right]},
\end{equation}
where $\mu$ and $\sigma$ represent the mean and standard deviation, respectively. In this study, we fix the minimum and maximum tour costs at $\pi/2$ and $3\pi/2$, respectively, generating other tour costs from the Gaussian distribution with $\mu = \pi$ and $\sigma = 0.5$, as shown in Table~\ref{tab:dataset}.

Fig.~\ref{fig:Grover_GM-QAOA}(a-c) illustrates the validation of the algorithm through numerical simulations of the quantum search among the total of $5! = 120$ feasible solutions for the 5-TSP problem. In this scenario, the non-solution states are concentrated near $\pi$, while the solution states are distributed around the tails of the Gaussian distribution. Consequently, the phase difference is approximately $\pi$, enabling the periodic evolution of success probability over time, as shown in Fig.~\ref{fig:Grover_GM-QAOA}(b). This behavior is similar to that of Grover's algorithm, leading to a quadratic speedup. 

Fig.~\ref{fig:Grover_GM-QAOA}(c) presents the probability distribution at $t = t_2 \sim 7$, with indices 1 and 120 corresponding to the solutions of the minimum and maximum costs, respectively. Since the quantum search amplifies both the minimum and maximum cost tours simultaneously, post-processing is required to extract the minimum cost tour~\cite{bang2012quantum}. Given that the computational complexity of this post-processing is $O(1)$, the overall computational complexity remains $O(\sqrt{n!})$.

\section{Numerical simulation of GM-QAOA}
\label{SecondAppendix}

The time evolution of quantum states by GM-QAOA~\cite{bartschi2020grover} is given by
\begin{equation}
\ket{\psi(\beta,\gamma)}=\prod_{i=1}^{p}\hat{U}_M(\beta_i)\hat{U}_P(\gamma_{i})\ket{\psi(0)}.
    \label{eq:GM-QAOA}
\end{equation}
Here, $\hat{U}_P(\gamma_i)=e^{-i\gamma_i \hat{H}_c}$ is the phase separator unitary which is diagonal in the computational basis.  $\hat{H}_c$ is cost Hamiltonian of TSP given by Eq.~(\ref{eq:obj_tsp}).   
We translate the classical cost function $\hat{H}_c$ into a quantum system by using Pauli-Z operators.  The relation between binary variable $x$ and Pauli-Z operators is given by
\begin{equation}
 x_{i,j} = \frac{1-\sigma_{t_s,i}^z}{2},
\end{equation}
where $\sigma_{t_s,i}^z \in \{-1,1\}$ is the Pauli-Z operator acting on the qubit corresponding to city $i$ at step $t_s$.  The Hamiltonian $\hat{H}_c$ is updated as 
\begin{align}
\hat{H}_c  &=  \sum_{i,j=1,i\neq j}^n \phi_{i,j}\sum_{t_s=1}^{n-1}\left( \frac{1-\sigma_{t_s,i}^z}{2} \right)\left( \frac{1-\sigma_{t_s+1,i}^z}{2} \right)\\
& = \sum_{i,j=1,i\neq j}^n\sum_{t_s=1}^{n-1}\frac{\sigma_{ij}}{4}(1-\sigma_{t_s,i}^z-\sigma_{t_s+1,j}^z + \sigma_{t_s,i}^z\sigma_{t_s+1,j}^z ).
\end{align}
Then we remove the first term and we obtain the unitary operator $\hat{U}_P(\gamma)$ as 
\begin{equation}
\hat{U}_p(\gamma) = \prod_{t_s=1}^{n-1}\prod_{i,j=1,i\neq j}^{n} e^{i\gamma\frac{\phi_{ij}}{4}(\sigma_{t_s,i}^z+\sigma_{t_s+1,j}^z)}e^{-i\gamma\frac{\phi_{ij}}{4}\sigma_{t_s,i}^z\sigma_{t_s+1,j}^z}.
\end{equation}
We can express the operator $\hat{U}_p$ with $R_Z, R_{ZZ}$ gates, such as  $e^{i\gamma\frac{\phi_{ij}}{4}(\sigma_{t_s,i}^z)} = R_Z^{(t_s,i)}\left( - \frac{\gamma\phi_{ij}}{2} \right)$ and $ e^{-\frac{i\gamma\phi_{ij}}{4}\sigma_{t_s,i}^z\sigma_{t_s+1,j}^z} $$= R_{ZZ}^{(t_s,i),(t_s+1,j)}\left( \frac{\gamma\phi_{ij}}{2} \right) $.
$R_{ZZ}^{(t_s,i),(t_s+1,j)}$ is a two-qubit gate applied between qubit  $(t_s, i)$ and qubit $(t_s+1, j)$.

$\hat{U}_M(\beta_i)=e^{-i\beta\ket{F}\bra{F}}$ is the mixing unitary.  $\ket{\psi(0)}$ is an equal superposition of all feasible solutions in $F$:
\begin{equation}
    \ket{F} = \hat{U}_s\ket{0} = \frac{1}{|F|}\sum_{x\in F}\ket{x}.
\end{equation}
Fig.~\ref{fig:Grover_GM-QAOA} (d) and (e) show the result of GM-QAOA for $n=4$ cities. We performed $500$ shots to calculate the expectation value $\braket{\psi(\beta,\gamma)|\hat{H}_c|\psi(\beta,\gamma)}$, and used the COBYLA optimizer provided by Qiskit for parameter optimization.

\section{Numerical simulation of SSQS}
\label{ThirdAppendix}
\begin{figure}
    \centering
    \includegraphics[width=5.0in]{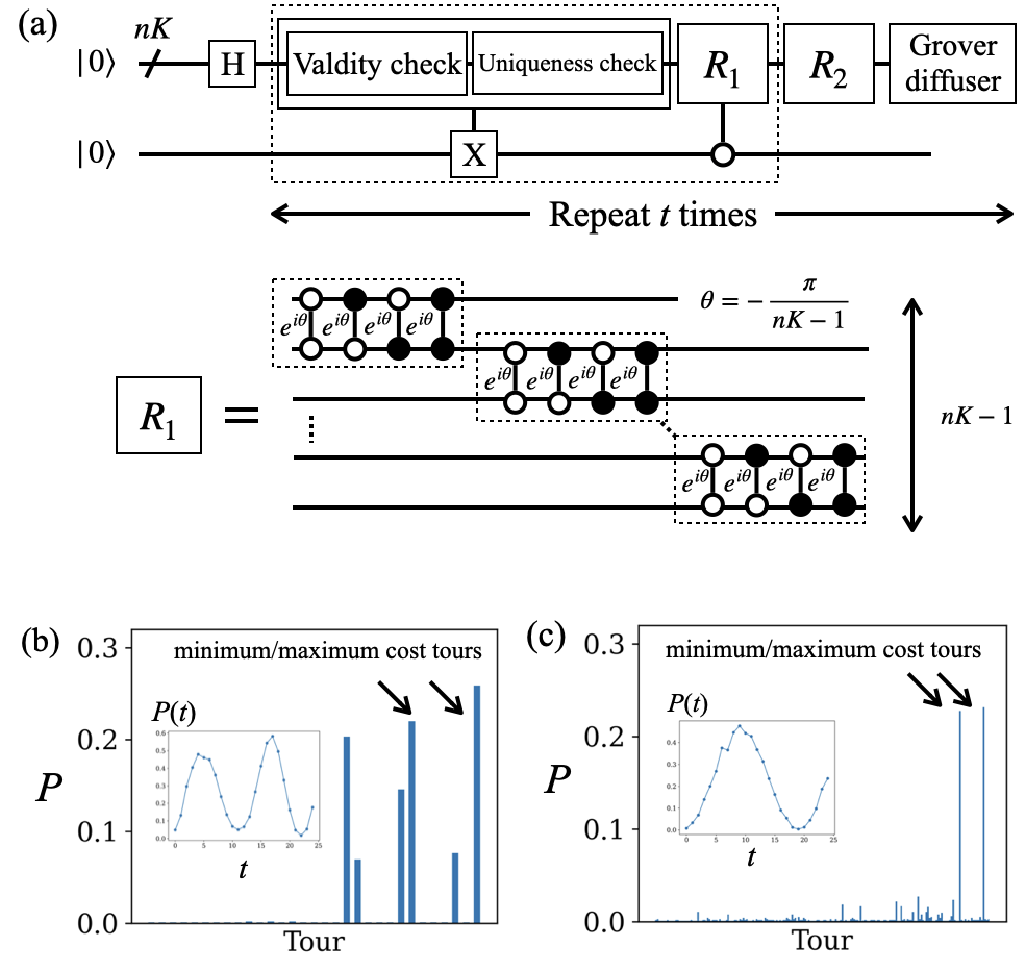}
    \caption{The simulation result of SSQS of solving TSP.  (a)The quantum circuit of SSQS.  The circuit of $\hat{R}_1$ is an example for $n=3, 4$. The white and black circle means $\ket{0}$ and $\ket{1}$, respectively.  (b)The simulation results of SSQS for $n=3$ TSP.  (c)The simulation results of SSQS for $n=4$ TSP.  The insets show the time-dependent success probability of the SSQS.\label{fig:result_SSQS}}
\end{figure}
The single-step quantum search (SSQS) algorithm evolves the system as
\begin{equation}
    \ket{\psi(t)} = (\hat{D}\hat{R})^{t} \ket{\psi(0)},
\end{equation}
where $\hat{R}$ is the oracle operator marking the minimum-cost tour state $\ket{T_{\min}}$, and $\hat{D}$ is the diffusion operator acting on an encoding space of dimension $2^{nK}$.  When the tour follows a Gaussian distribution, the tour states, except for the minimum and maximum tour states, are required to have a cost of approximately $\pi$, as shown in Fig.~\ref{fig:Grover_GM-QAOA}. Therefore, we design the oracle $\hat{R}$ using two sub-oracles, such that $\hat{R} = \hat{R}_2\hat{R}_1$, where $\hat{R}_1$ is applied to non-feasible states, which are given by 
\begin{equation}
\hat{R}_1 \ket{x} =
\left\{
\begin{array}{rl}
\ket{x} & \text{if } x = T_i,\\
e^{-i\pi}\ket{x} & \text{if } x \neq T_i,\\
\end{array}
\right.
\end{equation}
and $\hat{R}_2$ is applied to feasible state, given by
\begin{equation}
\hat{R}_2 \ket{x} =
\left\{
\begin{array}{rl}
e^{-iW(T_i)}\ket{x} & \text{if } x = T_i,\\
\ket{x} & \text{if } x \neq T_i.\\
\end{array}
\right.
\end{equation}
Fig.~\ref{fig:result_SSQS} shows the circuit of the SSQS algorithm for solving TSP of Table.~\ref{tab:dataset}.  To encode all states using an multi-controlled phase gate would result in a massive embedding cost given by $\mathcal{O}(2^{nK})$.  Therefore, we reduce the embedding cost by employing $nK - 1$ phase gates to encode all possible city states shown as Fig.~\ref{fig:result_SSQS} (a).  Additionally, the reason for setting $\theta = -\frac{\pi}{nK - 1}$ is to ensure that the final coefficient for any state is $\pi$.  

From the simulation results of Fig.~\ref{fig:result_SSQS} (b) and (c), we can evaluate the query complexity is $t_{\rm SSQS}=4, 9$ for $n=3$ and $4$, respectively.  The whole result is shown as Table.~\ref{tab:circuit_evalulation}.

\section*{Acknowledgment}
For part of this work, S.W. was supported by JST, PRESTO and JPMJPR211A.  S.W. thanks Shigeru Yamashita for his helpful comments.

\bibliographystyle{unsrt}
\bibliography{paper}

\end{document}